\documentclass[sigconf]{acmart}

\AtBeginDocument{%
  }

\copyrightyear{2026}
\acmYear{2026}
\setcopyright{cc}
\setcctype{by-nc-nd}
\acmConference[DIS '26]{Designing Interactive Systems Conference}{June 13--17, 2026}{Singapore, Singapore}
\acmBooktitle{Designing Interactive Systems Conference (DIS '26), June 13--17, 2026, Singapore, Singapore}
\acmDOI{10.1145/3800645.3812872}
\acmISBN{979-8-4007-2563-0/2026/06}

\usepackage{array}
\newcolumntype{L}[1]{>{\raggedright\let\newline\\\arraybackslash\hspace{0pt}}m{#1}}
\newcolumntype{C}[1]{>{\centering\let\newline\\\arraybackslash\hspace{0pt}}m{#1}}
\newcolumntype{R}[1]{>{\raggedleft\let\newline\\\arraybackslash\hspace{0pt}}m{#1}}
\usepackage{colortbl} 
\definecolor{mypink2}{RGB}{239, 240, 255}
\newcommand{\rowcol}{\rowcolor{mypink2}}
\newcommand{\rowwhi}{\rowcolor{white}}

\newcommand{\sys}{\textsc{UnReality Camera}}  
\begin{document}

\title{Exploring Instant Photography using Generative AI: A Design Probe with the UnReality Camera}


\author{Michael Yin}
\affiliation{
  \institution{University of British Columbia}
  \city{Vancouver}
  \state{BC}
  \country{Canada}
}
\email{jiyin@cs.ubc.ca}
\orcid{0000-0003-1164-5229}

\author{Angela Chiang}
\affiliation{
  \institution{University of British Columbia}
  \city{Vancouver}
  \state{BC}
  \country{Canada}
}
\email{achi2048@student.ubc.ca}
\orcid{0009-0008-8960-5837}

\author{Robert Xiao}
\affiliation{
  \institution{University of British Columbia}
  \city{Vancouver}
  \state{BC}
  \country{Canada} 
}
\email{brx@cs.ubc.ca}
\orcid{0000-0003-4306-8825}

\begin{teaserfigure}
\centering
  \includegraphics[width=0.95\textwidth]{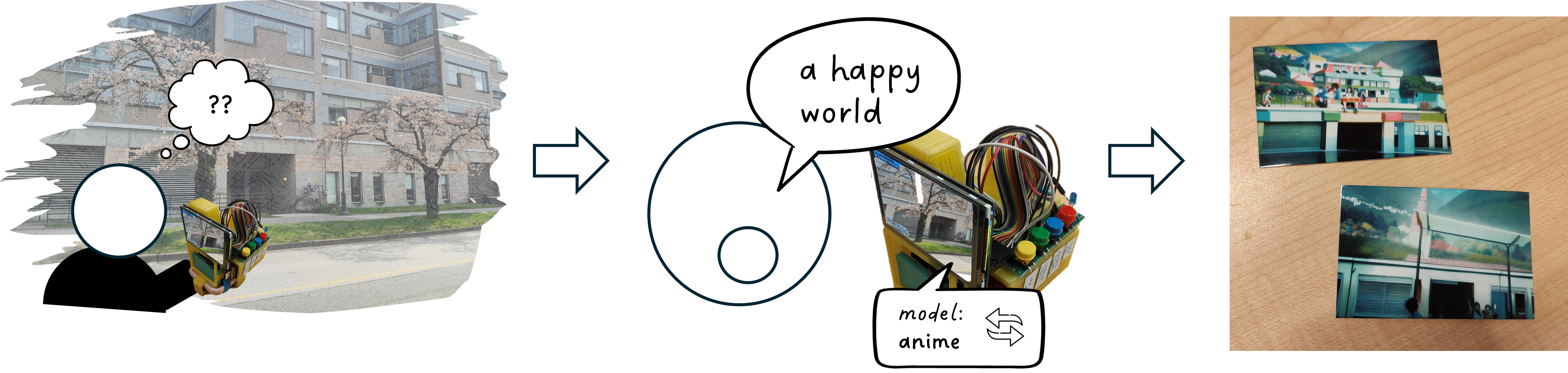}
  \caption{The \sys{} incorporates generative AI in instant photography. A person finds an environment that they would like to capture. While framing the shot, the person provides a spoken description and chooses a stylistic model that shapes the photo. The final image is printed out, representing an AI-mediated transformation of the captured environment. }
  \Description{The image is split into three subimages in flowchart form. In the first image, a person holding the UnReality Camera system is in a location they would like to capture. They are thinking about how they would like to capture it. In the second subimage, the person provides a spoken description, which is captured by the system (a happy world), and selects a generative AI model. Finally, the image prints out the generated images, which maintain qualities of both the original environmental surroundings and the spoken description mediated by the model.}
  \label{fig:teaser}
\end{teaserfigure}

\begin{abstract}
Generative AI has increasingly been used for artistic creation, but little work has explored how it shapes the \emph{experiential meaning} of practice. We consider how generative AI might transform the embodied and tangible process of instant photography through the \sys{}, an AI-mediated instant camera. The \sys{} prints a photo of the environment augmented by a user's spoken words as generative input. In a design probe, we explored how generative AI shapes people's perceptions of both photographic output and the broader photographic process. Although users valued artistic control, they also appreciated the creativity afforded by stochastic unpredictability. The waiting period for an unpredictable output elicited anticipatory suspense, and the camera's physical form evoked ownership and connection despite artificial generation. We discuss how people make sense of instant photography’s experiential qualities when generative AI is embedded, and how their opposing affordances reshape interpretations of each other's experiential meaning. 
\end{abstract}

\ccsdesc[500]{Human-centered computing~Empirical studies in interaction design}

\keywords{Generative AI, art, photography, camera, design probe, Polaroid}


\maketitle

\section{Introduction}

The artistic experience of \textbf{instant photography}, slow, tangible, and suspenseful, shaped its perception as a ``\emph{magical}'' process to capture a singular moment \cite{buseSurelyFadesAway2008a, murphyShakeItPolaroid2018} and fundamentally changed the way people experience materiality, temporality, and ease in photographic practices \cite{murphyShakeItPolaroid2018, busePolaroidDigitalTechnology2010a}. Photography has always been a unique form of creative expression, focused on capturing snapshots of reality \cite{edwardsObjectsAffectPhotography2012a}. Yet, while some perceive such snapshots to be objective and reproducible, their meaning is shaped by the photographer's intention, social context, and broader perception \cite{edwardsObjectsAffectPhotography2012a, skopikDigitalPhotographyTruth2003a}, in addition to the device and techniques used \cite{prakelComposition2020, battCameraCraftLearning2014, talInteriorLandscape2022}. Thus, photography is not only shaped by the output image, but the \emph{process} underlying it. 

\textbf{Generative AI} is another technology that sometimes shares the same moniker of ``\emph{magic}'' \cite{binderTechnologyDisEnchantmentAlphaGo2024b}. Generative AI systems have exploded in popularity, given their potential to write stories, help with research, support education, and, most relevantly, create visual ``\emph{art}'' through text-to-image or image-to-image models\footnote{\url{https://openai.com/index/dall-e-2}}\footnote{\url{https://www.midjourney.com/explore}}. Generative AI has been increasingly explored as a collaborative tool to support artists and extend creativity \cite{ohLeadYouHelp2018, shojaeiInsightsArtTherapists2024, wanItFeltHaving2024a, bryan-kinnsUsingIncongruousGenres2024, geroMetaphoriaAlgorithmicCompanion2019}. Artists may rely on such systems when they are stuck, taking advantage of the unpredictability of AI responses to spur inspiration \cite{wanItFeltHaving2024a, caramiauxExplorersUnknownPlanets2022}. They may also anthropomorphize these systems as \emph{facilitators} or \emph{subordinates} \cite{ohLeadYouHelp2018, shojaeiInsightsArtTherapists2024}, forming a relationship that affects their trust, perception, and usage of AI in their tasks. Yet, many artists are resistant to incorporating generative AI as well \cite{kawakamiImpactGenerativeAI2024a, jiangAIArtIts2023c, allredArtMachineValue2023a} due to the loss of authenticity, connection, and the human storytelling element \cite{messerCocreatingArtGenerative2024, allredArtMachineValue2023a, parkWeAreVisual2024}. 

We explore embedding generative AI into instant photography --- two processes that differ in \emph{temporal} (instant and re-generatable vs. delayed and deliberate) and \emph{material} (digital vs. physical) qualities --- and how this embedding shapes the \emph{experience} and \emph{feelings} during the photo-taking process. To ground this exploration, we constructed a prototype of an AI-supported, Polaroid-inspired device: the \sys{}. Unlike a conventional camera, users pass a spoken description of how they want to mediate the captured environment via a microphone, which is transformed into a generative AI prompt. This prompt is combined with the base photograph to form the resulting image, an \emph{AI-mediated vision} based on reality but moulded by the user's spoken words, which is printed out by the machine as a physical keepsake. Instead of capturing a representative snapshot of the world like a Polaroid would, or simply altering an image through generative AI, the \sys{} incorporates idiosyncratic parts of both. 

We used our system in a design probe \cite{gullota2017digital, wallace2013probe}. Through participants' exploration with, and conversations about, our system, we explored the questions of:

\begin{itemize}
    \item \textbf{RQ1}: How do people construct, interpret, and emotionally respond to generative images that blur the line between reality and non-reality in the process of instant photography?
    \item \textbf{RQ2}: How do people interpret the experiential dimensions --- such as temporality and material form --- of photo-taking when generative AI is embedded into the process of instant photography? 
\end{itemize}

\textbf{RQ1} relates to the user relationship with the visual generative output, while \textbf{RQ2} relates to the user interpretation of generative AI within the overall process. Through our exploratory probe, we spotlight the many tensions between the two processes that comprise the experience --- such as the motivation of capturing the moment versus generating something unreal, the authenticity of snapshots versus the artificiality of AI, the physicality of instant photography versus the digital nature of AI --- and we highlight how participants interpret these tensions. The experience of using the \sys{} incorporates two intertwined, yet distinct concepts: ``unreality'', referring to the quality of the image being non-existent in the real world, and ``stochasticity'', referring to the inherent randomness of generative AI that opens up unreality. The latter is the mechanism that supports the generation of unreal images: stochasticity supports the exploration of the many unrealities that exist. 

We highlight three main contributions of our work: (1) the design of the \sys{}, a Polaroid-inspired camera device that takes pictures of the world mediated by generative AI, (2) a design probe to understand interpretation of both process and outcome of instant photography when generative AI is embedded, and (3) synthesis of our findings into opportunity areas for future research and design regarding the implementation of generative AI into slow, embodied, artistic processes. We highlight how the design of an artistic process, including its temporal, physical, and embodied qualities, shapes how people interpret and understand the experience. We explore how deliberately mixing contradictory expectations and symbolism (similar to a counterfunctional design \cite{pierceCounterfunctionalThingsExploring2014a} or playful photography \cite{petersenDesigningPlayfulPhotography2009}) could extend the possibilities of artistic experimentation.

\section{Related Works}

\subsection{Perceptions Regarding Generative AI}

The rapid adoption of generative AI has become a critical, and sometimes controversial, issue in domains such as education and creativity. Generative AI has helped with efficiency and productivity, but has given rise to open questions regarding authenticity and trust \cite{wuReactingGenerativeAI2024a, allredArtMachineValue2023a}. In addressing these questions, we consider prior research on people's mental models regarding AI, which informs understanding how to build suitable systems \cite{khadpeConceptualMetaphorsImpact2020, binderTechnologyDisEnchantmentAlphaGo2024b}. Binder \cite{binderTechnologyDisEnchantmentAlphaGo2024b} highlighted a dichotomy between an \emph{enchantment} perspective --- viewing AI as a wondrous, almost magical concept --- versus a \emph{disenchantment} perspective --- perceiving AI as simply an ordinary tool. Kelley et al. \cite{kelleyExcitingUsefulWorrying2021} conducted a global survey regarding attitudes towards AI, and found that people grouped them into broad categories of the technology being ``exciting'' (representing hope and advancement), ``useful'' (representing benefit and assistance), ``worrying'' (representing caution and distrust), and ``futuristic'' (representing change). People have increasingly ascribed both positive and negative meanings to generative AI as being supportive tools but also potential human replacements \cite{grassiniHopeDoomAIttitude2023, laneToolTyrantGuiding2024}. 

The contextual perspective through which people view, think about, and discuss AI deeply affects how people trust and integrate AI into their daily lives \cite{binderTechnologyDisEnchantmentAlphaGo2024b}. Khadpe et al. \cite{khadpeConceptualMetaphorsImpact2020} considered how presenting different conceptual metaphors surrounding AI affected users' cooperation with the agent, their intention of use, and the usability of the system, shaping their overall relationship with the system. The relationship that people build with AI affects their attitudes and emotions towards themselves as well --- Kobiella et al. \cite{kobiellaIfMachineGood2024a} found that the use of LLMs could both enhance personal perceptions of productivity and output (e.g., when it acted more as a helper) and reduce senses of accomplishment, ownership, and self-adequacy (e.g. when it overshadowed their efforts). This can be mediated by constraints on interaction with the AI system, e.g. requiring a level of human time and effort needed \cite{joshiWritingAILowers2025}, or through design, e.g. through cognitive forcing functions such as a waiting period before an AI response \cite{bucincaTrustThinkCognitive2021}.

Oh et al. \cite{ohUnderstandingHowPeople2020a} found that people's attitudes toward understanding AI are broadly shaped by their life knowledge and domain of expertise. People use different techniques to better ``understand'' the AI, wanting to communicate with the AI for increased interpretability \cite{ohUnderstandingHowPeople2020a}. Van Brummelen et al. \cite{vanbrummelenAlexaCanProgram2021a} and Oh et al. \cite{ohLeadYouHelp2018} noted that interactions with AI agents can help people anthropomorphize them and assign human-like qualities, thus generating a familiar human-like relationship akin to a friend, helper, or subordinate. Extending on these human-like qualities, Wellner \cite{wellnerDigitalImaginationIhdes2022a} theorized on the question of whether AI has a human-like imagination. Overall, understanding people's perspectives towards AI affects how people use such systems in their processes and make meaning from its output. Our work uses the \sys{} as an exploratory probe into the interpretation around generative AI when it is embodied into an artistic process and physicalized as a process and an output. 

\subsection{Generative AI in Creative Domains}

HCI research has explored the use of generative AI in creative domains to support art and design \cite{yaoLuminaSoftwareTool2024a, gangulyShadowMagicDesigningHumanAI2024, shojaeiInsightsArtTherapists2024, bryan-kinnsUsingIncongruousGenres2024}. Artists assign roles to AI helpers, as a \emph{subordinate} who follows instructions \cite{ohLeadYouHelp2018}, a \emph{facilitator} that initiates conversation \cite{shojaeiInsightsArtTherapists2024}, or an \emph{audience} that evaluates and provides feedback \cite{bryan-kinnsUsingIncongruousGenres2024}. Beyond alleviating artists' workloads \cite{sunUnderstandingHumanAICollaboration2024a}, these roles enable AI to empower humans in exploring their thoughts \cite{shojaeiInsightsArtTherapists2024}, illuminating existing ideas \cite{wanItFeltHaving2024a}, and extending creative representation and thinking \cite{sunUnderstandingHumanAICollaboration2024a, kawakamiImpactGenerativeAI2024a, shojaeiInsightsArtTherapists2024, ohLeadYouHelp2018}. However, a common thread through almost all AI collaboration works in the creativity domain is that humans still want to be in charge of their creative output \cite{ohLeadYouHelp2018, johnstonUnderstandingVisualArtists2024, bryan-kinnsUsingIncongruousGenres2024, guoExploringImpactAI2024, sunUnderstandingHumanAICollaboration2024a, geroMetaphoriaAlgorithmicCompanion2019}. 

More specifically, Halperin et al. \cite{halperinUndergroundAICritical2025} categorized approaches to AI usage in filmmaking as minimization (minimally using AI), maximization (using AI as much as possible), compartmentalization (siloing its role), and revitalization (embracing serendipitous narratives). Tied to this last category, even though the lack of control over AI outcomes can be frustrating \cite{parkWeAreVisual2024}, their unexpectedness and unpredictability can spur surprise and inspiration in creative contexts. Caramiaux and Alaoui \cite{caramiauxExplorersUnknownPlanets2022} highlight how surprising results --- errors and glitches --- offer a sense of creative magic; Wan et al. \cite{wanItFeltHaving2024a} state that imperfect ideas and randomness introduce new concepts and remind users of creative possibilities. This inherent, abstract creativity is taken advantage of in the \emph{Dream Painter} exhibit \cite{guljajevaDreamPainterInteractive2022a} and the \emph{Dream Sticker Machine} \cite{bonlykkeTakingBizarreSeriously2024c}, both of which invite viewers to interpret visual AI representations of abstract dreams. 

However, art is the domain with possibly the greatest pushback against AI integration. Questions about legality regarding copyright and the ethics of ownership are important issues \cite{kawakamiImpactGenerativeAI2024a, jiangAIArtIts2023c, xuWhatMakesIt2024, johnstonUnderstandingVisualArtists2024}, yet Allred and Aragon \cite{allredArtMachineValue2023a} deem the most nebulous factor as the loss of authenticity, the connection between the art and its human creator. This connection comes from the expression of human experiences and storytelling \cite{parkWeAreVisual2024}, symbolic meaning which is lost with black-box AI agents \cite{shojaeiInsightsArtTherapists2024, sunUnderstandingHumanAICollaboration2024a}. Authenticity is considered important in artistic domains; art can be a way for humans to understand and engage in the world \cite{perkinsArtUnderstanding1988} or to express human feelings and themes \cite{langerCulturalImportanceArts1966, reidMeaningArts2014a}. Although AI models can discern such themes with subtle techniques, they can also misinterpret them \cite{halperinCameraEyeAIExploring2025}. Human-human collaboration takes advantage of social communication, dynamic adaptation, and shared contextual knowledge for mutual interpretation \cite{deshpandePerceptionsInteractionDynamics2024}; all of which are lost when interacting with generative AI. 

The affordances of generative AI offer both benefits and challenges in artistic domains. For instance, its ambiguity and unpredictability can both be a source of serendipity but also a barrier to usability or authenticity. Prior work has primarily focused on how these affordances shape how users collaborate with AI to shape their desired artistic outcomes. In our work, we highlight how these affordances --- stochasticity, ambiguity, unpredictability ---  become dimensions that shape the artistic \emph{experience} as well, i.e. the process of instant photography. 

\subsection{Photographic Tradition and Photography in HCI}

In the past, photography was described as an art form that focused on objective representations of reality \cite{laxtonMoholysDoubt2016}, highlighting the `truth claim' tied to photographic indexicality ---  the relationship between the photo and captured reality \cite{furuhataIndexicalitySymptomPhotography2009, gunning2004s}. Snyder and Allen \cite{snyderPhotographyVisionRepresentation1975a} stated that photographs, as representations of the real world, were inherently constrained by reality. Yet photographers, such as Moholy-Nagy, experimented with technique, perspective, temporality, and tools to create photographic works that depicted abstract and surreal images \cite{laxtonMoholysDoubt2016, newhallPhotographyMoholyNagy1941}. Techniques employing generative AI are the latest ventures in the lineage of experimental photography to capture something \emph{beyond reality}.

In the 1900s, Polaroids (i.e., instant photography) transformed the temporal experience of traditional photography, capturing a moment nearly instantly \cite{murphyShakeItPolaroid2018, buseSurelyFadesAway2008a}. Combined with its analog, material nature, instant photography was perceived to be more ``authentic'' \cite{minnitiPolaroid20PhotoObjects2016}. Murphy described such photography as being simultaneously analog, ``instantaneous'', everyday, and visual \cite{murphyShakeItPolaroid2018}. The shift re-shaped the ``temporal and phenomenological dimensions of taking, viewing, and sharing the analog visual experience'' \cite{murphyShakeItPolaroid2018}; for instance, people would shake the Polaroid print to hasten its resolution. Although Polaroids eventually fell out of fashion, they have been recently revived with a return to nostalgia \cite{lathropliguerosResistingObsolescencePolaroid2020, toyRiseDigitalReproduction2024} and a drive for ``anti-perfection'' \cite{lathropliguerosResistingObsolescencePolaroid2020}, in reclaiming autonomy as resistance against the turning of time. 

On the other end, modern smartphones have made photography accessible and convenient to the vast populace \cite{vanhousePersonalPhotographyDigital2011a}. As smartphones were easily transportable and connected to the Internet, photography became deeply intertwined with sharing the everyday \cite{liPlayfulnessImmediacySpontaneity2017, petersEverydayImageryUsers2018}, and such photographs became artifacts of instantaneous consumption \cite{liPlayfulnessImmediacySpontaneity2017}. Yet, as smartphones dominated the photography space, the physical photographic artifacts of the past that people used to hold and share have increasingly become lost and \emph{dematerialized} \cite{lathropliguerosResistingObsolescencePolaroid2020, minnitiBuyFilmNot2020}. The significance of photography is tied to the norms, culture, and social desires in which it exists \cite{edwardsObjectsAffectPhotography2012a}; with a shift in the process of photography based on social desires for efficiency and portability, intangible elements were both gained and lost. 

Prior research has probed the process of photography --- modulating the tools, temporal quality, and output --- to investigate how design can support new possibilities and critical discourse, both for photographs and broader design \cite{pierceCounterfunctionalThingsExploring2014a}. In Pierce et al.'s Counterfunctional Things \cite{pierceCounterfunctionalThingsExploring2014a}, design probes are presented (e.g. a camera where you cannot control when it takes a photo, or a camera where images cannot be viewed easily) that restrict the traditional affordances of a camera and recall past technologies. By deconstructing assumed wants, design probes open up new ways of thinking about and possibilities for existing technologies \cite{pierceCounterfunctionalThingsExploring2014a, seoBack1990sBeeperRedux2025b, odomDesigningSlownessAnticipation2014}. Odom's \emph{PhotoBox} \cite{odomDesigningSlownessAnticipation2014} randomly prints old photos, supporting discourse on domestic technology and how slowing down technology can support reflection and anticipation. Boucher et al.'s \emph{TaskCams} \cite{boucherTaskCamDesigningTesting2018} show short texts that ask questions or request images, becoming useful tools for cultural probes. Benjamin et al.'s \emph{Entoptic Field Camera} \cite{benjaminEntopticFieldCamera2023} generates synthetic outputs for user inputs based on their smartphone images, providing insight into human understanding of reality. Karmann's \emph{Paragraphica}\footnote{\url{https://bjoernkarmann.dk/project/paragraphica}} is a context-to-image camera to use location data with generative AI to visualize an imagined location, eliciting moods and emotions based on AI's understanding and perception of place. Most similar to our work, Page et al.'s \emph{A(I)Cam} \cite{pageCreativeReflectionsImageMaking2025} facilitates AI-created images as a tool for creative inspiration and exploration, re-shaping how users see and understand the world. 

Page et al. \cite{pageCreativeReflectionsImageMaking2025} (and to a lesser extent, Benjamin et al. \cite{benjaminEntopticFieldCamera2023}) overlap our work in terms of design and perception of photographic outcome, as their work also considers the experience of generative photography through design probes. However, our system differs in terms of functional design, i.e. having a way for the user to pass a prompt to mediate the captured image through voice, which reshapes user expectations of usage and their relationship to the process and the output. By having a text-based mediating input, our system falls more similarly to a non-diegetic input, with Page et al.'s work being more diegetic \cite{dangChoiceControlHow2023}. Furthermore, our work uniquely considers how the affordances of AI shape not just perception and understanding of the visual output, but of the entirety of the \emph{process of instant photography}, including its human-centric, material, and temporal qualities.

\begin{figure*}[h]
  \centering
  \includegraphics[width=0.9\linewidth]{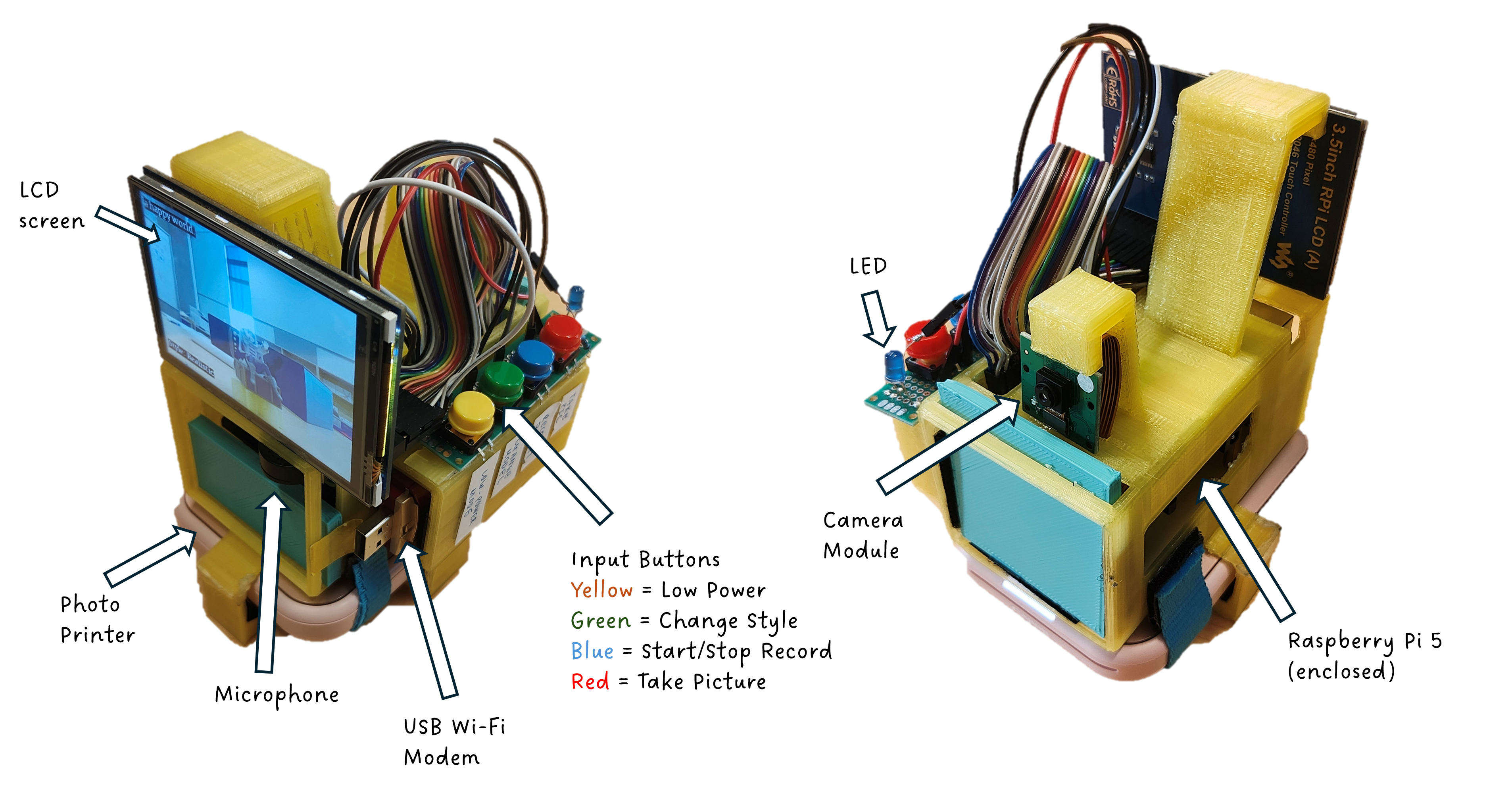}
  \caption{Front and back view of the \sys{}. The USB Wi-Fi Modem is presently unplugged, but can be connected to the Raspberry Pi for a portable Internet connection.}
  \Description{Images taken from the front and back views of the system. From the front view, we see the LCD screen that displays the environmental surroundings, the photo printer that prints out the generated images, the Wi-Fi modem that can be used for a portable Internet connection, and the input buttons that the user interfaces with to change models, record spoken descriptions, and take images. From the back, we see the LED that is used for feedback and the camera module that captures the environmental surroundings. The Raspberry Pi 5 is enclosed within the device.}
  \label{fig:system}
\end{figure*}

\section{System Design}

The \sys{} is a portable prototype of a deconstructed camera that incorporates generative AI into its resultant images. Users can walk around with the camera and point it at the surroundings they would like to capture, which is reflected on the system's display. Users can also provide a spoken-word description of how they would like the captured reality to be transformed. The system processes this description, blending it with the real photo to create a generated image, which is then printed on photographic sticker paper. Here, we discuss how our research questions led to the development of the system, and how our design decisions help probe into its experience and use.  

\subsection{Design Goals for a Design Probe}

As a design probe, the \sys{} aims to provoke discussion regarding our research questions \cite{wallace2013probe}. Similar to other design probes \cite{odomDesigningSlownessAnticipation2014, pierceCounterfunctionalThingsExploring2014a, gullota2017digital}, our goal was to elicit understanding of these questions through provocation rather than to necessarily optimize for usability; human involvement during the design stage could potentially risk reducing the frictions in human-AI collaboration that we inherently set out to explore. Recalling our research questions, \textbf{RQ1} focuses on people's perceptions regarding the output of the image, blurring the line between what people expect photography to capture (i.e. the real world) versus a potentially unreal, imagined output. \textbf{RQ2} concerns people's perceptions regarding the experience when generative processes are embedded into instant photography. Thus, the design of the \sys{} was informed by principles drawn from these research questions.

\begin{itemize}
    \item \textbf{Subverting the Expectations of Outcome} - We sought to contrast typical photography's consistent and faithful representation of the environment against our system's unpredictable output. In our system, the user never truly knows what the output image may look like, even with a mental model of expectations; later in the paper, we explore this role of unexpectedness. Although any sort of stochastic mediation may have sufficed, we chose to use generative AI due to its contemporary relevance. Thus, our system also offers potential insights and discourse into generative AI's well-studied effects, as well as relative controversy, within artistic processes.
    \item \textbf{Evoking an Expressive, Yet Familiar Experience} - We wanted to recall instant photography's actual physical, temporal, and embodied experience --- of carrying a device, walking around, and waiting for the image to print. Regarding \emph{physicality}, we considered the importance of the tangible form of the device \cite{jungDigitalFormMateriality2012} as something that the users can hold, use to frame the world, and click. Furthermore, physical form also plays into the capture of a physical one-shot image. Tying into \emph{temporality}, there was an inevitable waiting time during the generation and printing process, and this contrasted against the ephemerality of a singular captured moment. Finally, the \emph{embodied} nature of photography, being able to take the system around and freely capture the world at any geographic location, was replicated as well through the implementation of the system. 
\end{itemize}

Creating the physical device (rather than building a smartphone app that can also induce a delay or create printouts similar to instant photography) was important because the experience of instant photography is inevitably entangled with its material form. Smartphones (and their photos) entail specific symbolic meanings corresponding to instantaneity, consumption, and abundance \cite{louEffectsMobileIdentity2022, chopra-gantPicturesItDidnt2016, liPlayfulnessImmediacySpontaneity2017} that are inseparable from the material form, which ultimately shapes the experience of photographic practice. 

We built the \sys{} to deliberately evoke the feeling of using an instant camera, aiming to induce nostalgia through usage that can reveal possibilities for design \cite{seoBack1990sBeeperRedux2025b, pierceCounterfunctionalThingsExploring2014a}. We aimed to emulate the experience and \emph{feel} of instant photography through its design (e.g. as in BeeperRedux \cite{seoBack1990sBeeperRedux2025b}), yet integrate the modern technology of generative AI to transform its \emph{outcome}. The physical object and the physical print are important in recalling the symbolic value of instant photography --- as a nostalgic, slow, and authentic process that resists digital change \cite{lathropliguerosResistingObsolescencePolaroid2020}. We deliberately integrate this with the contrasting symbolism of AI --- as efficient and future-facing, representing excitement, worry, and change \cite{kelleyExcitingUsefulWorrying2021}; in contrast, a smartphone app would have symbolic value that is more strongly aligned with generative AI. 

One key connecting question was how exactly the generative AI system should affect the captured reality. Highlighted now and discussed later, this question gives rise to a new task of guiding the system's understanding of the imagined intention. For the \sys{}, we decided to allow users to pass simple voice input that lets users \emph{tell} the system how they wanted the image to appear. This contrasts with Page et al.'s \emph{A(I)Cam} \cite{pageCreativeReflectionsImageMaking2025}, as we provide a level of control for users to modulate the captured world. We chose spoken voice because it aligns more strongly with the existing flow of the photographic process, unlike text input, which would require the user to disengage with the framing and capturing of the shot. Thus, this helps maintain the familiar actions involved in instant photography, aligning with the design principles. Furthermore, spoken words can be more spontaneous and unfiltered compared to written text \cite{sawhneyAudioJournalingSelfreflection2018a, yin2026reflective}, helping to capture instantaneous reactions to the environment. 

\subsection{System Implementation}

In this section, we expand on the technical implementation of the \sys{}. Given the prototypical nature of the system for this probe, we built the \sys{} as a modular and cohesive system integrating existing and low-cost components. 

The base computational unit of the system is a Raspberry Pi 5. A 5-megapixel video camera module captures the system's surrounding environment, which is displayed on a 3.5-inch LCD screen. Users can provide spoken word descriptions using a mini USB microphone, which the system processes to generate prompts. The system requires a stable internet connection to interface with the AI elements. In our studies, we used an existing Wi-Fi connection due to its relative stability; however, the system also has an option to house a portable cellular modem. This, combined with two 18650 Li-ion rechargeable batteries for power, allows the system to be used portably. 

Finally, to print out the stickers, the system interfaces with a Canon Ivy 2 Mini Photo Printer via Bluetooth, leveraging an open-source Python API\footnote{\url{https://github.com/dtgreene/ivy2}}. The user interacts with the system using four buttons on its side, and feedback is displayed either as text on the LCD screen or through an LED light component. A 3D-printed enclosure houses and protects all the components, maintaining a compact, portable, and holdable form (see Figure \ref{fig:system}).

\subsection{System Flow}

\begin{figure*}[h]
  \centering
  \includegraphics[width=0.8\linewidth]{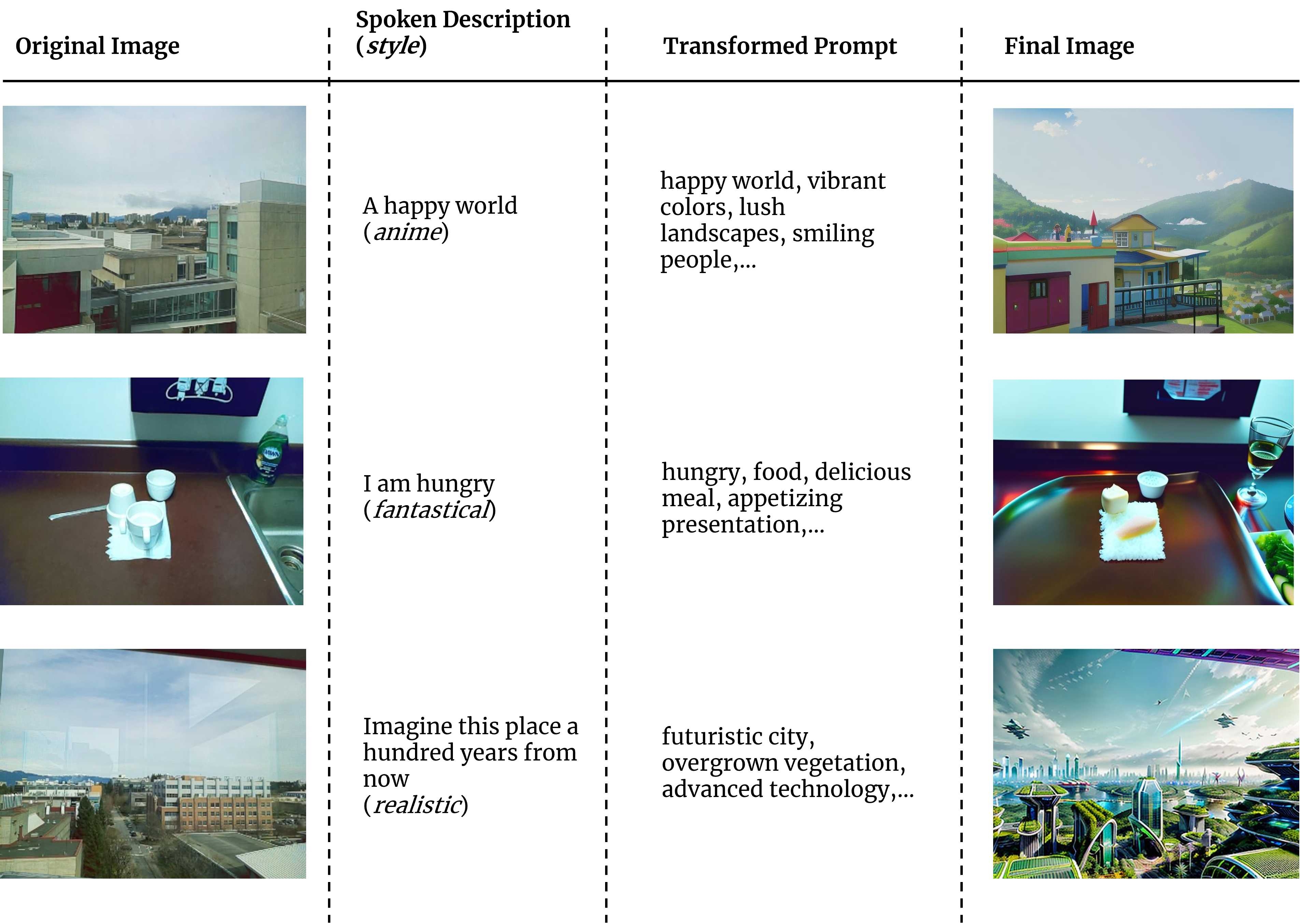}
  \caption{Some examples of the original and final images using the \sys{}. A user's spoken description is first transformed into a comma-separated prompt. The combination of the prompts with the models (styles) and the original images results in the final images on the right.}
  \Description{This image tabulates the original image, the spoken description, the style, the transformed prompt, and the final image. The first row takes an image of several concrete buildings in front of mountains with the description "a happy world" and the anime styling, and the resulting image shows a colourful, idyllic environment in the mountains. The second row takes an image of some dirty cups, dishwasher soap, and a sink with the description "I am hungry" and a fantastical styling, and the resulting image shows cake, a wine glass with a drink, and salad. The third row takes an image of some buildings along a road with the description "Imagine this place a hundred years from now" with the realistic styling, and the resulting image shows a futuristic city.}
  \label{fig:table}
\end{figure*}

\subsubsection{Capturing and Processing Input}

The system takes two input channels: visual and auditory. For visual input, the video camera module automatically captured the surroundings, which were then lightly processed (e.g. adjusting resolution) before being displayed on the LCD. For auditory input, users pressed a button to start and stop recording to capture audio data. The audio input transcription was displayed on the LCD for user reference. The audio data was transcribed using OpenAI's whisper-1 model, and subsequently translated into an image generation prompt using OpenAI's GPT-4o-mini model, with the prompt \emph{``provide ONLY a comma-separated prompt with image generation tags for stable diffusion for the input''} and the role \emph{``You help me write prompts for Stable Diffusion''}. This process helped translate raw spoken language, which could include filler and ambiguous wording, into a structured keyword-based syntax which aligns with training. Keyword-based prompts are widely adopted in the generative AI community to create more controllable outputs \cite{oppenlaenderTaxonomyPromptModifiers2024a} akin to prompt expansion techniques \cite{dattaPromptExpansionAdaptive2024}. We tested this informally during implementation; however, we acknowledge that this adds a level of interpretation to the user's intention. 

\begin{table*}[h]
  \caption{Participant demographic information and self-reported experience with photography (on a scale of Beginner / Intermediate / Advanced / Expert) and generative AI (on a scale of None / Limited / Moderate / Extensive)}
  \label{tab:demographics}
  \begin{tabular}{c|c|c|c|c}
    \toprule
    Participant ID & Age & Gender & Photography Experience & Generative AI Experience \\
    \midrule
    \rowcol P1 & 28 & Man & Intermediate & Extensive \\
    \rowwhi P2 & 27 & Man & Beginner & Moderate \\
    \rowcol P3 & 32 & Man & Beginner & Moderate \\
    \rowwhi P4 & 24 & Man & Advanced & Extensive \\
    \rowcol P5 & 27 & Woman & Intermediate & Extensive \\
    \rowwhi P6 & 22 & Woman & Intermediate & Moderate \\
    \rowcol P7 & 27 & Woman & Intermediate & Limited \\ 
    \rowwhi P8 & 28 & Man & Intermediate & Moderate \\
    \rowcol P9 & 23 & Man & Intermediate & Extensive \\
    \rowwhi P10 & 22 & Man & Intermediate & Moderate \\ 
    \rowcol P11 & 22 & Woman & Intermediate & Moderate \\
    \rowwhi P12 & 22 & Woman & Advanced & Moderate \\
    \rowcol P13 & 19 & Woman & Advanced & Extensive \\
    \rowwhi P14 & 18 & Man & Advanced & Moderate \\
    \rowcol P15 & 48 & Woman & Advanced & Limited \\
  \bottomrule
\end{tabular}
  \Description{Table displaying participant information collected during the design probes. Attributes include age, gender, self-reported experience with photography, and self-reported experience with generative AI.}
\end{table*}

\subsubsection{AI Image Generation}

We chose to use the generative AI model of \emph{Stable Diffusion v1.5}, providing a blend of speed and quality, while also being accessible in terms of price, locally runnable, and configurable by the researcher. Although this specific model was chosen for our system, in essence, any sort of stochastic generative model would have been adequate for our probe. To interface with the generation model, we used the API provided by the popular Stable Diffusion Web UI\footnote{\url{https://github.com/AUTOMATIC1111/stable-diffusion-webui}}, which ran locally on a computer (Intel Core i7-10700K processor and NVIDIA GeForce RTX 3080 graphics card) and was exposed using ngrok\footnote{\url{https://ngrok.com/}}. In particular, we used the \textbf{img2img} endpoint, which takes an image (of the camera-captured surroundings) and a payload (with the image generation prompt) to generate a new image. 

To introduce diverse possibilities for image generation, we added four fine-tuned variants of Stable Diffusion v1.5 to create different artistic styles. We assigned them each a one-word descriptor --- \emph{Anything V3} (anime)\footnote{\url{https://civitai.com/models/66/anything-v3}}, \emph{DreamShaper V8} (whimsical)\footnote{\url{https://civitai.com/models/4384/dreamshaper}}, \emph{OpenJourney V4} (fantastical)\footnote{\url{https://civitai.com/models/86/openjourney}}, and \emph{Realistic Vision V6.0 B1} (realistic)\footnote{\url{https://civitai.com/models/4201/realistic-vision-v60-b1}}. Users could cycle through different fine-tuned styles for the image generation process, and the currently selected style was displayed on the LCD with its one-word descriptor. 

The parameters for the image generation, such as the \emph{number of steps} (20), \emph{denoising strength} (0.55), and \emph{cfg\_scale} (20), were subjectively tuned through trial and error until the values seemed adequate for creating an image that largely retained qualities of the captured surroundings but was still different; see Figure \ref{fig:table} for examples. Given the unpredictable nature of AI-generated images, we also incorporated negative prompting to filter out NSFW and harmful content. 

\subsubsection{Printing and Physical Output}

The final generated image was printed as a physical output by the Ivy 2 printer, connected to the Raspberry Pi through Bluetooth. This served as a tangible artifact for the users to interpret and use as they see fit. As our system takes time to perform all the processing steps, the printing does not start instantaneously after the button click. Instead, an informal test on 10 printed snapshots indicated that the printer would begin printing (i.e. an audible printing sound) at an average of approximately 21 seconds after clicking, and the entire print could be collected after approximately 63 seconds. This is comparable, if not faster, to the speeds of typical Polaroid cameras \cite{busePolaroidDigitalTechnology2010a}.

\section{Exploratory Design Probe}

We invited participants to use our system in a controlled setting to situate their experience and interpretations of mixing generative AI with instant photography. Through people's experience with this provocative system, we aimed to reveal insights into how this hybrid practice might reframe people's thoughts about instant photographic processes, reactions to the unreal images, and imagined futures for incorporating generative AI into embodied, artistic processes. Importantly, we focussed on using the \sys{} to spark conversation and discussion, rather than evaluating the system as a tool. While we considered deploying the system in longer field studies (and we recognize this as important future work), we chose to perform the study in a more controlled lab setting to (1) offer instant assistance and discussion with the users if needed, and (2) elicit instantaneous responses from initial usage, beyond practical reasons such as time and cost. 

\subsection{Participant Recruitment}

Participants were recruited through a mix of convenience sampling and a listing made on our institute's paid studies postings. The eligibility criteria were to be 18 or older and to be able to use our system to take pictures in and around our institute; we also purposively sampled for a range of photographic experiences. We recruited a sample of 15 participants (age ranging from 18 to 48, with a mean of 25.9; 8 reported as men, 7 as women). In addition to these demographic checks, we also queried for their self-reported experiences with photography (4-point Likert scale --- beginner / intermediate / advanced / expert) and generative AI (4-point Likert scale --- none / limited / moderate / extensive) --- see Table \ref{tab:demographics} for a more complete overview. Before the session, we asked participants to read, review, and sign a consent form outlining this research's ethics and data usage; ethics approval was obtained from our institute's ethics review board. 

\subsection{Study Protocol}

Upon arrival, participants were provided with an introduction to the overarching research and reviewed a consent form regarding data collection and usage. The researcher then demonstrated the functions of the \sys{}, walking the participant through the various functions. 

After the introduction, participants were given the hands-on opportunity to use the system to generate photos. We intentionally kept the tasks open to treat exploration itself as important --- we simply asked participants to freely explore the local area in and around our institute and take images of different environments as they wished (although we did ask them to exclude identifiable photos of other people, for privacy reasons and respect for third-party ethics). Although we encouraged them to try different prompts and models, participants were mainly left to explore freely. The openness was an intended feature of this study; we wanted users, given a system, to try to construct their own ideas of the system, their own experiments with using it, and, in doing so, identify possible motivations and scenarios for future interaction. We accompanied the participant as they explored to provide technical support and converse about their thoughts and actions with the \sys{}, although we acknowledge that the presence of the researcher may have shaped participant use and experience (e.g., increased self-consciousness).

After a period of around 50 minutes, we moved into the final semi-structured interview, where the researcher asked the participants about their experience with using the system, their emotions, interpretations, and meaning-making processes (see supplemental material for an outline of the general questions). Participants reflected on the images they had taken, which ranged in number from 7 -- 10 (average: 8.7). Given the qualitative, abstract nature of our research questions, we felt that an open-ended interview was the most appropriate method; through these conversations, we were able to surface the latent implications regarding the unpredictability of generative AI mediation, and their relationship to the physical, temporal, and embodied process of instant photography. Each session lasted approximately 80 -- 90 minutes, and participants were compensated \$24 CAD for their participation. 

\subsection{Data Analysis}

We analyzed the qualitative data through a thematic analysis approach \cite{braun2021TA}. As our interview was largely framed around existing research questions, we followed a mainly deductive paradigm. After familiarizing themselves with the data, the lead researcher coded the interview scripts with a focus on the experiential content. This process began with a semantic coding of the qualitative data, which was iteratively refined and adjusted during the process. The codes were then iteratively mapped into broader categories through an affinity diagramming approach (see supplemental materials), which formed the themes that informed our findings around our research questions. The lead researcher mainly performed this data analysis approach, but the findings were discussed with the research group to gather other perspectives. We highlight that our approach is exploratory, looking at the breadth of various ways in which AI mediation might shape people's photography experiences rather than focusing on quantification or generalizability. We interpret our data to highlight insights around the relationships between generative AI, humans, and photography.

\section{Findings}

We use the resulting themes from our data analysis to address our initial research questions, painting vignettes of how participants engaged with and interpreted their experiences. 

\subsection{Developing and Responding to Generative Output}

\emph{\textbf{RQ1}: How do people construct, interpret, and emotionally respond to generative images that blur the line between reality and non-reality in the process of instant photography?} 

\subsubsection{Constructing Desired Outcomes through Negotiated Authorship}
\label{sec:negotiation}

Participants adopted strategies to navigate the stochastic, yet rule-bound space of prompting, while lacking a full understanding of how the prompting and generation process worked. When participants had a specific unreal scene that they wanted to capture in a certain imagined way, their way of constructing it was through verbal communication with the device. This process, unique to our system compared to other instant photography systems \cite{pageCreativeReflectionsImageMaking2025}, became a point of communicative friction. For some participants, trial-and-error processes were used to negotiate their intended goal with the algorithm: 

\begin{quote}
    \emph{``I think I sort of struggled to like, verbalize what I wanted to say, and I fell back to these like one or two word prompts that were very short and nonspecific.''} - P10
\end{quote}

This highlights one of the key difficulties of human-AI co-creation --- the alignment of intention, especially through words as the medium for nuanced intention communication. Participants resorted to trial-and-error to adjust their prompts to align with their desired outputs. As P13 stated: \emph{``I'm looking at how I can change my input to get the result that I want''} (P13). To construct the images they wanted, participants engaged in a \emph{dialogue} with an AI system, one that often could not perfectly understand what they wanted (either due to missing context clues, lack of environmental understanding, etc). This dialogue caused friction in the layers of authorship that would not typically exist in traditional instant photography, due to misalignment at potentially several steps --- the actual human intention, their linguistic expression of their intention, the algorithmic interpretation of the words, and the generated image based on that interpretation. This friction in authorship was also exacerbated by temporal friction. P2 and P10 mention the \emph{``iterative''} process of artistic creation, which is lost in the time it takes to \emph{``wait [for] it to be printed''} (P2).

\subsubsection{Interpretation of the Artificial Generation}
\label{section:interpretation}

Participants' experiences, including their expectations and interpretations of the unreal images, were dependent on their shifting perceptions and understandings of generative AI within the task. Participants analogized the AI as having a wide variety of different roles relative to themselves, e.g., as an \emph{unsupervised intern} (P1), an \emph{employee} (P3), a \emph{bad assistant} (P5), a \emph{director} (P7, P11), a \emph{collaborator} (P12) or \emph{co-creator} (P2), and so forth. These roles were established based on the interrelated concepts of expectations, control, and perceived authorship, and were reminiscent of those found in prior works on generative AI systems \cite{ohLeadYouHelp2018, shojaeiInsightsArtTherapists2024, panchanadikarImSoloDeveloper2024, vanesYourFriendlyAI2025}.

Participants interpreted the hierarchy of roles between themselves and the AI based on how they perceived the balance of work in creating an output. Many of the participants initially expected themselves to maintain the \emph{``main role of the photographer''} (P8), with the AI acting as an \emph{``assistant to help me shape what I want to show''} (P8). Yet, participants ended up having varied perceptions regarding their input. Some participants thought they had a high degree of input: \emph{``I am the creator because I was the one who said the model and the prompt''} (P5); others felt like they had minimal input: \emph{``My role is simply prompting and the AI did everything else''} (P3). This was subjective, as the person's procedural involvement was constant --- they pointed the device, chose a model, gave a prompt, and clicked the button to take an image. Relatedly, participants' interpreted roles were tied to their perceived level of control. For instance, if the AI was perceived to have more of a role in the final photo, then the participant may feel \emph{``I was the assistant''} (P12).  

Participants' assignments of the AI's role also depended on their orientation towards the AI --- whether it was to build a specific vision or to openly interpret an idea. Again, this varied largely participant to participant --- some participants had strong expectations on what the output should be, e.g., \emph{``add things into the photos''} (P3), \emph{``tweak up the image''} (P4); others had loose initial expectations, e.g., \emph{``I didn't have really any picture in my head about what it would have looked like, I didn't really have any expectation about it''} (P13). The expectation that the participant envisioned for the output image affected their perception of it as helpful or disruptive. With a looser set of expectations, participants could backfill the generative AI's interpretation of their prompt:

\begin{quote}
    \emph{``For this one, where [I prompted] `I'm tired, heading home', it interpreted the kind of staircase, plus the walls on the side, plus the railing, as a subway train home.''} - P14
\end{quote}

\subsubsection{Emotional Response and Desires}
\label{section:emotion}

Given the context of participants' construction and interpretation of the generative AI imagery, participants experienced and expressed a wide range of different emotional responses towards the final unreal image. At one extreme, participants felt adverse feelings such as disappointment and vexation --- e.g. \emph{``frustration, irritation, ... lead someone who's feeling that way to not use it''} (P11), and \emph{``it could be frustrating that someone wanted a specific thing''} (P12). These feelings arose when the user had a specific imagined output that they wanted:

\begin{quote}
    \emph{``[if] it does meet your expectations and if you enjoy AI, it's satisfying. In this case, I found like, if it generates completely different than what I wanted to, it's a bit off-putting.''} - P14
\end{quote}

When the participants had a stronger sense of their imagined output --- a more goal-focussed interpretation --- they would respond positively when AI met these expectations (e.g. for P2, when the photos \emph{``I [felt] are going the exact direction that I expected, in this case I'm happy''}), and negatively otherwise (e.g. for P3, the images were \emph{``jarring''} because they \emph{``completely changed the background... I felt like I was just generating AI images''}). Stochastic generation fundamentally creates a chasm between human intentions and algorithmic interpretation, an inherent gap in human-AI co-creation, as subjective interpretations may never align perfectly. When humans approached their use of the \sys{} with a strong intention and expectation, they often reacted negatively to interpretation: \emph{``in general it's negative because rarely do I have something in mind and then it is better than what I have imagined''} (P3).  

Yet, when the participants had a looser, open interpretation of their imagined output --- a more exploratory interpretation --- they were more likely to respond positively. In this case, it was less about matching their vision but using the generative AI to more freely interpret a concept, with an open interest in what it would come up with. For instance:

\begin{quote}
\emph{``The prompt was `I am hungry', which I didn't have like expectations about the outcome. For these cases, I kind of think that it was more fun and serendipitous.''} - P5
\end{quote}

P13 found that it was more interesting than negative to \emph{``see what an AI brain interprets as what I said''}, and P15 stated that \emph{``it didn't bother me that they all turned out different''} indicating that \emph{``the surprise factor is there... they're all creepy in their own way''} (note: creepy was used positively here, as a sense of eerie delight). P8 stated that, although the prompt indicates what they want to see, they did not expect the AI to have priors and just wanted the AI to provide \textbf{its} interpretation, indicating that \emph{``because you never know the outcome, you can see some very interesting or shocking stuff''}. From this perspective, unpredictable stochasticity becomes a feature that potentially serves better to explore and inspire rather than create a specific vision. All in all, participants' emotional responses hinged on not only the actual output image, but also their approach, interpretation, and orientation towards the system and towards uncertainty. 

\subsection{Embedding Generative AI in the Experience of Instant Photography}

\emph{\textbf{RQ2}: How do people interpret the experiential dimensions of photo-taking when generative AI is embedded into the process of instant photography? } 

\subsubsection{Tangibility and Personalization}

As foregrounded by participant responses already, art is as dependent on the creation process as much as the outcome. We revisit the effects of the tangible, physical dimension of both the photo-taking system and the resultant output image. Firstly, the physical form of the system offered a sense of fun and whimsy, as P3 mentioned that this caused them to approach it with a \emph{``whimsical and candid nature of a Polaroid camera versus trying to take the best picture possible''}, and P1 stated that it felt more \emph{``like a toy rather than a software''}. The sense of having a weighty, material system that they could hold provided users with a rather indescribable sense of value, personal expressiveness, and intentionality:

\begin{quote}
    \emph{``Personally, I love the more physical interaction that holding a `camera' provides, and the tactility of the buttons... significantly different than holding a phone. Fresh, novel, more liberating, whereas the phone feels kinda like a `sink', no responsiveness and very 1-dimensional.''} - P11
\end{quote}

\begin{quote}
    \emph{``I personally think holding a physical, camera-shaped object makes me feel like the photo is more intentional... however, taking photos on a phone feels more casual. If the study was done on a phone, I feel like the sense of control over the result would be even less than having the study done with the camera-box.''} - P12
\end{quote}

Participants indicated that physicalizing both the output and the device made the process feel more intentional, more proximate, and more significant when compared to the ubiquity of a phone. For P10, \emph{``a physical camera is more meaningful''} and for P14 \emph{``the AI is in your hands right now... It's less significant when you consider it on your phone''}. Part of this significance arises from having a device with a sole purpose of taking images, as P10 stated that with such a device, they are \emph{``less likely to get distracted and `taken out of the moment' ''}; P5 extended that \emph{``holding the object that is specifically designed for photography makes me feel more engaged, because when I take pictures using a phone it's just one of many functions''}. The physicality of an intentional device offered a way to recapture meaning, significance, and resonance with the specific practice. 

The physical nature of the \emph{printed output} transformed participant responses towards the outcome of photography. A physical output heightened the sense of ownership and belonging that the participants felt in co-creating the image, e.g. \emph{``printing out the pictures makes me feel stronger ownership''} (P5). Participants stated that: 

\begin{quote}
    \emph{``Having like a physical representation of what you took is more... it's not just using the visual sense. It's also like you're holding it and you can see it and share it in ways that aren't strictly just visual.''} - P13
\end{quote}

\begin{quote}
    \emph{``You just get this instant physical thing in your hands that in some ways has more value to it than just a little digital photo on your phone's camera roll.''} - P10
\end{quote}

Yet, this sense of ownership comes in tension with the inherent loss of feelings of control and authorship that comes in AI-generated art in the first place, outlined in \ref{section:interpretation}. We find that physicality helps offset the loss of control or ownership that comes with generative AI-assistance and transforms stochasticity into a form of personal uniqueness. It transforms the generated image into almost a random reward, or as P2 coins it, a \emph{``collectible''}; one where \emph{``there's only one copy of that photo in the entire world, and you can't really replicate it''} (P12). 

\subsubsection{Temporality and Stochastic Friction}

Another important dimension of instant photography is temporality. The first temporal period we consider is the waiting period between the user's click of the button and the actual printed output being received. Initially, we considered this waiting period an inconvenient inevitability, as filler time for the generative algorithm to produce an output and for the data to be sent and printed. Yet, we observed this waiting period to become an important part of the photo-taking experience as a form of \textbf{``stochastic friction''} --- which we define as the anticipatory, suspenseful waiting period where people hope, wonder about, and speculate about their ``reward'', of what the outcome could be. As highlighted in \ref{section:emotion}, the randomness of the output meant that people's reactions can range from surprise and inspiration to disappointment; feelings in this waiting period built up to those emotions:

\begin{quote}
    \emph{``It’s almost like a one-time interaction; you provided something, and you are waiting for feedback. So in that moment, I do have some expectation and curiosity in seeing the result.''} - P2
\end{quote}

\begin{quote}
    \emph{``I guess I am expecting... imagining what the result the system might give. A sense of anticipation before the actual printout.''} - P8
\end{quote}

We also consider the temporal dimension of the machine in the broader context of human desires. For participants, the rather instantaneous waiting time offers a way to ``try again'' if the first image did not come out the way that they imagined --- this was a practice that we observed for some participants, who took multiple pictures, perhaps with a slightly tuned prompt, to get something closer to what they had in mind. Yet, the waiting time of the device also made retries difficult and effortful. This offers a very candid, one-shot approach to capturing the moment similar to instant photography, over which you have minimal control over the single output image --- P7 indicates that this instantaneity allows one to \emph{``try and capture something in the moment without so much planning behind that''}, P11 analogizes the appeal as \emph{``being in the moment and you don't have a second chance''}. P15 finally states that for Polaroids, you never knew what you were going to get, drawing a parallel to the appeal of our system in precisely that \emph{``it's a crapshoot''}. This inherent randomness introduces an aesthetic and an emotional risk, when it fails (e.g. \emph{``By the time you get [the output], you're not there anymore''} - P7), the moment is lost. 


Our findings underscore tension regarding the temporal element of generative stochastic AI. As AI systems aim for increased speed and efficiency, they also strip away parts of anticipation (through friction) and momentary significance (through effort), which contribute to the ``\emph{wonder}'' of technologies such as instant photography.

\subsubsection{Situated Expressiveness and Human Interaction with the \sys{}}

Putting it all together, the process of using the \sys{} was human-focused and situated. Instead of making it easy for the user to imagine and create their desired output, it requires them to carry a device, walk around, wait, and explore. We recall participants walking around and exploring different places, such as the institute's rooms and hallways, the grassy walking paths outdoors, and even empty bathroom stalls. Participants tried many different framed shots and a variety of prompts, such as trying to add a porcupine into the grass, or imagining the buildings a hundred years from now (some examples are found in the supplemental materials). Even when the results were sometimes disappointing, participants contrasted with a much easier and tunable process of prompting a model:

\begin{quote}
    \emph{``I think even if I sat for like an hour and prompted a model and like created the exact thing that I was wanting to, I think I would have trouble feeling like I created this content just because I didn't go through the process of actually making it.''} - P10
\end{quote}

Calling forth some of the participants' definitions of art, we again highlight the importance of the process beyond the outcome --- P11 mentioned that two similar images made in different ways are \emph{``inherently very different''}, highlighting how the \textbf{process} of creation reframes perception. P5 stated that \emph{``art is creating something fun with a purpose... in terms of that, I think this is art''}, and P14 states that \emph{``it's more personalized to you because you took the effort and time to [make] that one photo''}. This contrasts with the controversy of ownership that comes from AI usage in art, which removes the human authenticity from creative endeavours. This tension was evident in other contrasting participant responses --- \emph{``I don't consider [it] art because I don't think I had creative direction or control''} (P11). Synthesizing everything, our findings underscore the intrinsic value of the experience of instant photography. Time, effort, and human involvement have increasingly become abstracted away, yet these underscore feelings of wonder, anticipation, ownership, and resonance.
    
\section{Discussion}

We considered how our findings are contextualized around existing views of human-AI collaboration in creative domains and how they fit into the temporal and physical contexts of design. Even when the actual process of taking an instant photograph remains constant, we highlight the experiential and interpretive responses when we intertwine generative AI with the process of instant photography. We also investigate the meanings attributed to experiential dimensions of temporality and material form. While instant photography and generative AI have been studied individually, we consider how placing opposing affordances in juxtaposition reshapes each other's experiential meaning. Based on these findings, we highlight actionable design considerations for the future. 

\subsection{Control, Expectations, and Perceptions of Generative AI in Instant Photography}

\subsubsection{Generative AI and Aligning to User Expectations}

Our findings that participants generally felt a loss of control compared to traditional photography agree with prior work on generative AI in the arts. Prior research highlights the tradeoff in autonomy for collaborative support, and that participants have generally wanted to be primarily in control \cite{ohLeadYouHelp2018, johnstonUnderstandingVisualArtists2024, bryan-kinnsUsingIncongruousGenres2024, guoExploringImpactAI2024, sunUnderstandingHumanAICollaboration2024a}. The sources of outcome control in our work were mixed. Similar to the \emph{A(I)Cam} \cite{pageCreativeReflectionsImageMaking2025}, our camera takes human input from the environmental capture and voice input, yet the final step of combining these two human inputs into a visual image is delegated to an unpredictable and stochastic AI. However, even though the system outcome was unpredictable, its behaviour was stable across participants. Instead, we find that participant \emph{responses} and \emph{feelings} towards the outcome were led through primarily individual perceptions of the roles of generative AI, the strength of their expectations, and their alignment towards the goal of taking such a photo. 

When we combine the expectations shaped through a relatively high degree of control (e.g. through framing the shot and passing in a prompt) with the inherent lack of control of generative AI, the resulting co-creation is often misaligned with what users want. People's strong expectations were informed by prompting, where they essentially explicitly told the AI what to do. Such expectations contrast systems with more abstract outcomes (e.g. the \emph{Dream Sticker Machine} \cite{bonlykkeTakingBizarreSeriously2024c}) or systems with weaker human control (e.g., the \emph{A(I)Cam} \cite{pageCreativeReflectionsImageMaking2025}). Although the prompting mechanism allowed for exploration of possibilities (aligning with \cite{dangChoiceControlHow2023}), it could also create a stronger alignment of expectations, which affects the judgment of agency \cite{didionWhoDidIt2024}. Our findings suggest that \textbf{increased input control created more expectations and consequently more frustration when expectations were not met reliably}. In contrast, we hypothesize that less control, like in A(I)Cam \cite{pageCreativeReflectionsImageMaking2025}, may align better with stochastic, abstract, and unpredictable outcomes. This potentially ties to psychological ownership as well --- increased time and effort expenditure (i.e. through prompts) results in the person feeling like they ``own'' the generated outcome \cite{joshiWritingAILowers2025}, perhaps as an expression of their control and identity \cite{pierceTheoryPsychologicalOwnership2001}. 

One short-term way of addressing the expectation gap is to improve the reliability and performance of the AI. Challenges in aligning intent in human-AI co-creation are presently widespread \cite{dangHowPromptOpportunities2022}. We draw inspiration from prior research on how the design of prompting systems can be improved to better align user expectations. Having multiple output options could offer more chances to meet expectations \cite{dangChoiceControlHow2023}, and being able to rapidly make incremental, reversible edits \cite{massonDirectGPTDirectManipulation2024}, e.g. through localized inputs \cite{gholami2023diffusionbrushlatentdiffusion}, would help users ensure the outcome meets their vision. Interactive elements and incorporating human decision-making can be useful for creating satisfying results that align with the user's creative vision \cite{huhVideoDiffHumanAIVideo2025, geroMetaphoriaAlgorithmicCompanion2019}. These features would help the system outcomes meet the expectations implied by our design \cite{palaniDontWantFeel2022}. 

\subsubsection{Generative AI and Serendipitous Experiences}

While our study revealed challenges that arose from strong user expectations, it also revealed that a positive experience --- serendipity --- could arise when user expectations were reframed. Unexpected outcomes and imperfect ideas can spur surprise, innovation, and creativity in the arts and beyond \cite{caramiauxExplorersUnknownPlanets2022, wanItFeltHaving2024a, dangWorldSmithIterativeExpressive2023, chenAutoSparkSupportingAutomobile2024}; in a photography case, this agrees with Page et al. \cite{pageCreativeReflectionsImageMaking2025}. In our study, this often required the users not to have a strong intention of the output. We found this to be a subjective element based on perception more than actual control, since participant involvement was largely constant, yet their interpretations and level of expectations differed; such expectations could potentially be shaped through past experiences, cognitive bias, and individual personality \cite{riedlTrustArtificialIntelligence2022, nouraniImportanceUserBackgrounds2022}. Thus, when expectations were misaligned with outcomes, it led to frustration; when expectations were loosened, it led to possible discovery. Expectations could be observed through the actual prompts that participants used throughout the studies, ranging from more targeted, objective, prompts with more concrete expectations (e.g. \emph{``add a porcupine here''} - P3), to more abstract ones (e.g. \emph{``I am hungry''} - P5) 

Stochastic generation of imagined realities shifts the expectation of reflecting reality into reflecting possibilities, transitioning from a photographic experience of \emph{documenting} reality to \emph{speculating} on it. Much like past experimentation with photography, such shifts directly impact the paradigm of photography \cite{laxtonMoholysDoubt2016, newhallPhotographyMoholyNagy1941} and its indexicality. We interpret capturing generative realities as entailing a different set of motivations, use cases, expectations, and behaviours when compared to capturing reality; thus, a prerequisite to discovery-based usage is to communicate to the user to set their expectations appropriately. For LLMs, this may include communicating early that it hallucinates \cite{metzeBibliographicResearchChatGPT2024}; analogized for our system, it may be communicating that the model might not always follow your instructions exactly. This could also be accomplished more abstractly --- such as mentioning that the system is meant to imagine possibilities rather than make specific edits. Through communication, the variety of roles that were assigned to the AI might become more distilled, reducing the confusion and frustration potentially generated from AI's ever-changing roles \cite{zhangProtoDreamerMixedprototypeTool2024}.

Aligning with previous work, AI-supported serendipity can induce feelings of inspiration and new possibilities \cite{ohLeadYouHelp2018, wanItFeltHaving2024a}, potentially helping pull artists out of mental blocks \cite{wanItFeltHaving2024a}. We find that such stochastic surprise also forms a unique source of anticipatory excitement and engagement (similar to a gacha pull \cite{yinRewardLuckUnderstanding2022}), where participants feel a range of emotions towards a randomized result. 

\subsection{Form and Temporality: Generative AI and the Process of Instant Photography}

\subsubsection{Physical Form and Psychological Ownership}

We revisit the form of our system, considering its visual and physical appearance \cite{jungDigitalFormMateriality2012}, as material form is important to symbolism meaning of the experience. Fuchsberger et al. \cite{fuchsbergerMaterialsMaterialityMedia2013} discussed how the materiality of a system impresses humans and interplays with human activities and experiences. The form of our device, reminiscent of a Polaroid, captures facets of expression and meaning \cite{jungDigitalFormMateriality2012}. We consider how simply having a tangible physical form enabled feelings of meaningfulness, significance, and ownership over the experience and artifact in our study. 

These feelings may partially stem from the symbolic value of material form --- participants differentiated from a phone, which comes with specific connotations \cite{chopra-gantPicturesItDidnt2016, louEffectsMobileIdentity2022}. Smartphones are multipurpose and ubiquitous, the \sys{}, in contrast, became purposeful and intentional. Furthermore, the anthropomorphism of the generative AI component of the camera worked on top of human tendencies to anthropomorphize physical objects in general \cite{boyerWhatMakesAnthropomorphism1996, wanAnthropomorphismObjectAttachment2021}, making the perception of the camera much different compared to a different photo-taking device, such as a phone --- they ascribed it anthropomorphic roles and felt a connection to the physical device. Connecting anthropomorphism to AI, Rozendaal et al. \cite{rozendaalGivingFormSmart2018} discussed the form of AI-incorporated objects in particular, discussing intelligence as character and considering their familiarity, authenticity, and sociability; strongly tying to our study's depiction of AI as playing certain roles. Aligned with prior work, participants described the AI using anthropomorphic language and assigned the AI agent in the camera with various personality traits \cite{sallesAnthropomorphismAI2020}. 

With this reframing in symbolic understanding and cognitive perception of the photographic object, participants adjusted their feelings to the process accordingly. We tie this to psychological ownership, as having a physical device could imbue a stronger personal value beyond the more digital smartphone \cite{atasoyDigitalGoodsAre2018}; the custom-built device being unique adds a sense of scarcity as well. When we juxtapose this against the less ``authentic'' process of generative AI, which entails a much more nebulous definition of, and feelings regarding ownership \cite{xuWhatMakesIt2024}, this harks towards how physical interfaces can paradoxically increase human ownership or connection even toward something as detached, uncontrollable, and non-interpretable as generative AI. 

The principle of physical ownership extends not only towards the physical camera, but also the physical output as a printout. When these outputs require human time and effort, imbueing their thoughts and feelings, and are scarce (i.e. limited by printing paper), they become artifacts that participants felt a sense of psychological ownership towards \cite{atasoyDigitalGoodsAre2018, pierceStatePsychologicalOwnership2003, pierceTheoryPsychologicalOwnership2001}. Embedded in generative AI, the physical materials support people in feeling more connected to the materials of the experience, taking stronger ownership of the output, and ascribing significance to the process even when action remains largely constant. Thus, while physicalizing a digital system for use in an embodied process requires more time and effort, its mixture with a stochastic process can \textbf{reframe the process to add individualization and ownership}.

\subsubsection{Temporal Delay and Wonder}

We also consider the temporal dimension of our system. The shift from viewing design as revolving around ``things'' to ``events'' has underscored the importance of time, which can affect how people reflect and derive meaning from their experiences \cite{wibergTimeTemporalityHCI2021}. Much work has studied the concept of slow technology, which offers a paradigm to design reflective and amplified experiences \cite{hallnasSlowTechnologyDesigning2001a}. Although instant photography came with the promise of rapid results, it still took time for the photographic result to become visible. This temporal dimension is mimicked by the \sys{}, which takes a comparable amount of time due to its generation and printing speed. We found that this waiting period, i.e. ``stochastic friction'', was a boon for building anticipation, similar to Odom et al.'s \emph{Photobox} \cite{odomDesigningSlownessAnticipation2014}, helping build anticipation towards an outcome and induce reflection \cite{odomDesigningSlownessAnticipation2014}. 

Similar to Photobox \cite{odomDesigningSlownessAnticipation2014}, which also incorporates a degree of known randomness, the stochasticity of generative AI in our work exacerbated anticipation. Participants, not knowing what the output might be, used this time to imagine and speculate. Yet, this contrasts with existing AI systems that prioritize speed. This delay also made it difficult for people to ``redo'' the image when it turned out in a way that they did not like, contrasting existing AI systems that allow for re-generation. The \sys{} introduces slowness into the process, which inhibits usability but establishes a sense of significance and uniqueness. During waiting times, people end up thinking \cite{bucincaTrustThinkCognitive2021, coxDesignFrictionsMindful2016a, tepponenUnderstandingWaitingVideo2025}, disrupting mindless use \cite{coxDesignFrictionsMindful2016a} and shifting into a System 2 paradigm under dual-system theory \cite{kannengiesserDesignThinkingFast2019}. For objective rather than creative tasks, this thinking can materialize as analyzing for the ``correct'' answer \cite{bucincaTrustThinkCognitive2021}. On the other hand, imbuing stochasticity into instant photography, a creative, abstract process with no ``correct'' answer, translates this thinking instead into anticipation and wonder. Waiting times can be detriments to usability and efficiency of generative AI tools and can exacerbate already existing issues with negotiating authorship (Section \ref{sec:negotiation}). Thus, while slowness in an artistic process can create frustration and inefficiencies, deliberate incorporation of slowness can also \textbf{recover the magic and wonder through anticipation}.

\subsubsection{Transforming the Meaning of Photographic Experience}

Altogether, the \sys{} integrates generative AI into the embodied, slow process of instant photography to probe into how it impacts the experience. Calling back to prior design probes, e.g. \cite{pierceCounterfunctionalThingsExploring2014a, seoBack1990sBeeperRedux2025b}, such exploratory probes are important because they show what is gained or lost through paradigm shifts --- changes in material, speed, and human involvement. 

By imbuing generative AI in a slow and tangible interface, we ultimately sacrifice some level of usability and ubiquity - the process can no longer be quickly performed, easily iterated on (if the output is disliked), and digitally processed without the need for a paper output. However, by making the system slow, deliberate, and physical, we imbue the process with a sense of personal significance and uniqueness --- a sense that personal time and effort were spent to create this expressive output. Even if the visual output was not necessarily enjoyed, we show that generative AI also plays a role in shaping the experience of creating this output, which is important as well. This agrees with how such factors can improve the \emph{aesthetics of deliberation} and the \emph{quality of action} from Seo et al.'s work \cite{seoBack1990sBeeperRedux2025b}. Altogether, we highlight that changes in temporal and physical qualities fundamentally change the goals of how people perceive, engage with, and \emph{wonder} about photography. Designers can leverage changes in form and time within stochastic practices as a way to trade off between an anticipatory and significant experience against a completely usable one. 

As users frame the unique moment, they capture a moment that is difficult to capture again. The experience becomes a performative \cite{jacucciInteractionPerformancePerformative2015} experience to express something personal and unique, \emph{even when} the outcome of such expression is ambiguous in tone and subjective in meaning due to embedded generative AI. In essence, the \sys{} is ironic. It materializes instant photography, a nostalgic process intertwined with human autonomy and rebellion against digital modernity \cite{lathropliguerosResistingObsolescencePolaroid2020, minnitiBuyFilmNot2020}, while incorporating generative AI, a contemporary, digital process that complicates feelings of human agency \cite{allredArtMachineValue2023a, ohLeadYouHelp2018, halperinAISoullessHollywood2025}. By fusing these, the \sys{} re-enchants generative AI, a technology often stripped of its magic by its ubiquity and efficiency, by giving it a physical form, a unique material output, and deliberate temporality. At the same time, it re-explores instant photography as not just a site for one-shot capture and material output, but a source of speculation, unpredictability, and wonder. 

Our discussion has explored how the stochasticity of the process and its integration within the temporal and physical experience of instant photography shape perception and interpretation of both the process and the outcome. We believe that this embedding can also fundamentally shift the basis of photographic meaning, tied to our initial concept of ``unreality''. By subverting traditional photographic indexicality, the \sys{} opens up new potentials of speculating on the different abstract ``unreal'' possibilities --- participants discussed how using the \sys{} enabled \textbf{speculation on \emph{what could be}, rather than \emph{reflection on what is}}, acting akin to a Monte Carlo simulation of reality. 

Changes in artistic process and meaning can hark towards shifting movements and cultural zeitgeists, fundamentally shifting how people produce and interpret art. For instance, Dadaism took advantage of chaos and randomness in artistic \emph{processes} to represent existentialist themes \cite{dzhimovaCalculatedRandomnessControl2024, rosenInvokingMuseDadas2014, kristiansenWhatDada1968}. Thus, artistic value and meaning arrive both from its outcome and the process of creation. While the \sys{} departs from traditional indexicality, the incorporation of stochastic randomness (similar to Dadaism) and temporal delay in both \emph{process} and \emph{outcome} can potentially open up a slower, reflective layer of understanding, meaning-making, and interpretation, fundamentally reshaping \emph{meaning} driven by \textbf{the design of the process}.  

\subsection{Ethical Acknowledgement}

We situate our findings around the broader social perception of and discourse around generative AI. AI art can be fun, amusing, and can serve as a social critique or speculation on imagined possibilities \cite{halperinUndergroundAICritical2025}. It also raises concerns when it undermines labour, takes over authentic expression, or defies people's wishes \cite{kawakamiImpactGenerativeAI2024a, jiangAIArtIts2023c, halperinAISoullessHollywood2025}. We probe into the embedding of stochasticity of generative AI in creative processes, but also strongly urge designers and researchers to critically reflect on the social and ethical implications of such technologies. For instance, our findings demonstrate that collaboration with an AI to create a visual image and printing it out as a physical artifact can elicit feelings of psychological ownership. Yet, this may be potentially problematic if a person feels like they solely ``own'' or ``created'' the generated image if it undermines authorship or erases the labour of creativity. 

\section{Limitations}
Limitations arose due to the controlled lab setting of our study, as participants were encouraged to take photos for exploration. This differs from contexts in which participants would naturally take photos, which may be more sporadic in timing and more significant in the captured environment. When we take the view that photographs are for capturing the moment for future remembrance, then participants sometimes indicated a need for more \emph{time} to process their photos and reflect on them. Future work could extend towards a full, longitudinal in-the-wild study in which the system is incorporated naturally into one's photographic routine over a longer period. Studying the extended use of the \sys{} could also help minimize the novelty effects from first-time use. This could also support a more detailed examination of the actual photographic artifacts themselves, including how users might perceive and experiment with different types of spoken prompts and the different styles. The perception of different generative styles could be examined in more detail as a factor of authorship and control; this factor was underexplored in our findings. 

Furthermore, participants were from a similar demographic range, and few had advanced experience with photography. While we found this adequate for an initial speculative exploration, we highlight how the use of photography as an art form rather than simply capturing the moment might tie more towards professional photographers; this could be an interesting population to purposively sample in the future, as we hypothesize that professionals might be able to offer a more detailed account of the latter motivation, as well as the potential for stronger baseline comparison against their existing practice (especially if they have extensive instant photography experience). While we did not stratify our data analysis based on photography experience, our informal observation of the data showed that people with advanced photography experience were more articulate at expressing their expectations, but still showed a range of responses from being frustrated at misalignment to being open to serendipitous interpretation. Future work could extend to a broader demographic, incorporating quantifiable work with statistical measures and causal inferences based on the impact of generative AI across different experiences. This could involve examining the frequency and valence of responses to provide more targeted takeaways.

Our work uses a specific prototype in a probing study to address broader questions regarding people's relationships with AI, perceptions of stochastic generation in creative domains, and the impact of experiential dimensions such as temporality and materiality. However, the \sys{} deliberately constrained certain factors of freedom and control, leading to an unexplored design space. Based on our foundation in understanding design, alternative designs could shift their focus to more deeply addressing usability of our system through, e.g. the customizability options desired by many participants that may impact alignment in co-creation (having the option to quickly iterate on an image before printing, being able to generate multiple images, incorporating memory of user preferences over time, etc.). Accessibility and the context of use are other explorations that would be valuable. For instance, the use of speech as prompt input may not work for all users (e.g. due to privacy or during situational impairment \cite{saulynasUnderstandingSeverelyConstraining2016}), and future work can look into alternative inputs and their effects on co-creation. 

\section{Conclusion}
We performed a design probe using the \sys{}, a Polaroid-inspired device that mixes instant snapshot photography with generative AI. The device does this through blending a user's spoken input with the environmental capture and a generative art model to produce and print out an output image, capturing the world as perhaps the AI might interpret it based on the user's words. From our probe, we found that users view the photographic process as co-creation with the AI, with the latter having roles based on their perceived level of control and the ability of the AI to meet their expectations. The inherent sacrifice of control in our system can negatively impact the user in trying to create their specific vision, but can induce creativity and serendipity when users approach it with more relaxed expectations. Even when the actions of taking an instant photograph remain largely constant, we consider how intertwining generative AI in instant photography reshapes how people perceive and interpret materiality and ownership, temporality and waiting, and the overall goal and \emph{meaning} of both processes. 

\begin{acks}

This work was supported in part by the Natural Science and Engineering Research Council of Canada (NSERC) under Discovery Grant RGPIN-2019-05624. 

\paragraph{Generative AI Disclosure} Beyond generative AI's actual integration into our system, generative AI was used in two ways in the project: (i) to help debug and aid in system programming, and (ii) to help scope down and discuss the broad research ideas (as a thinking partner). The writing was done by humans, with light AI assistance for specific word choices or grammatical improvements. 
\end{acks}
\bibliographystyle{ACM-Reference-Format}
\bibliography{sample-base}


\end{document}